\documentclass[pra,twocolumn,showpacs,groupedaddress,superscriptaddress,aps,10pt]{revtex4-1}
\usepackage{bm,graphicx,amsmath}
\usepackage{placeins}
\usepackage{amssymb}
\usepackage{amsmath}
\usepackage{epsfig}
\usepackage{amssymb}
\usepackage{color}
\usepackage[colorlinks,linkcolor=blue,citecolor=blue]{hyperref}
\usepackage{subfigure}

\begin{document}

\title{Type I and type II second harmonic generation \\ of conically refracted beams}
\date{\today}

\author{A. Turpin}
\affiliation{Departament de F\'isica, Universitat Aut\`onoma de Barcelona, Bellaterra, E-08193, Spain}
\author{Yu. V. Loiko}
\affiliation{Departament de F\'isica, Universitat Aut\`onoma de Barcelona, Bellaterra, E-08193, Spain}
\affiliation{Aston Institute of Photonic Technologies, School of Engineering \& Applied Science Aston University, Birmingham, B4 7ET, UK}
\author{T. K. Kalkandjiev}
\affiliation{Departament de F\'isica, Universitat Aut\`onoma de Barcelona, Bellaterra, E-08193, Spain}
\affiliation{Conerefringent Optics SL, Avda. Cubelles 28, Vilanova i la Geltr\'u, E-08800, Spain}
\author{J. Trull}
\affiliation{JDepartament de F\'isica i Enginyeria Nuclear, Universitat Polit\`ecnica de Catalunya, Rambla Sant Nebridi 22, E-08222 Terrassa, Spain }
\author{C. Cojocaru}
\affiliation{JDepartament de F\'isica i Enginyeria Nuclear, Universitat Polit\`ecnica de Catalunya, Rambla Sant Nebridi 22, E-08222 Terrassa, Spain }
\author{J. Mompart}
\affiliation{Departament de F\'isica, Universitat Aut\`onoma de Barcelona, Bellaterra, E-08193, Spain}

\begin{abstract} 
Type I and type II second harmonic generation of a beam transformed by the conical refraction phenomenon are presented. We show that for type I, the second harmonic intensity pattern is a light ring with a point of null intensity, while for type II the light ring possesses two dark regions. 
Taking into account the different two-photon processes involved in second harmonic generation, we have derived analytical expressions for the resulting transverse intensity patters that are in good agreement with the experimental data. 
Finally, we have investigated the spatial evolution of the second harmonic signals showing that they behave as conically refracted beams. \\
\textbf{ocis}:(190.2620) Harmonic generation and mixing, (160.1190) Anisotropic optical materials; (260.1180) Crystal optics; (260.1140) Birefringence. 
\end{abstract}

\date{\today}
\maketitle

\begin{figure}[ht!]
\centerline{\includegraphics[width=0.90 \columnwidth]{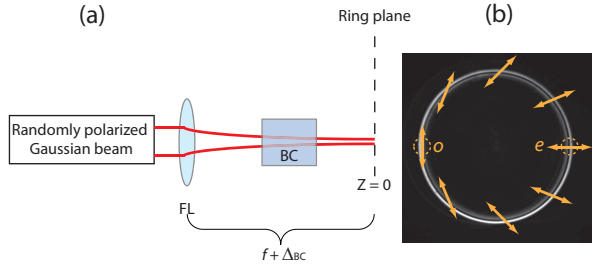}}
\caption{(a) A focused randomly polarized Gaussian beam is transformed by a BC into a light ring at the ring plane of the system. (b) CR ring at the ring plane with the fine Poggendorff splitting. Double orange arrows show the polarization distribution along the ring. FL means focusing lens. \textit{o} and \textit{e} denote the points with \textit{ordinary} and \textit{extraordinary} polarizations, respectively. $\Delta_{\rm{BC}}=L(1-1/n_{\rm{BC}})$ is a longitudinal shift of the ring plane added by the BC, with intermediate refractive index $n_{BC}$.}
\label{fig1}
\end{figure}

\begin{figure}[]
\centerline{\includegraphics[width=1 \columnwidth]{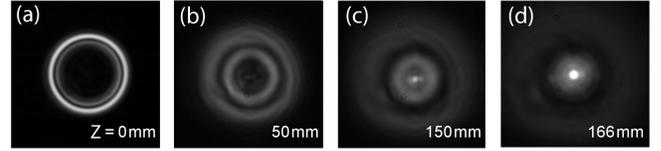}}
\caption{Evolution of the transverse intensity profile of the FH generated throughout the CR effect in a BC. The position of the ring plane, where the CR ring is most sharply resolved, is at $Z = 0$. Experimental parameters: $R_0=476~\rm{\mu m}$, $w_0=42~\rm{\mu m}$ and $z_R=5.148~\rm{mm}$}
\label{fig2}
\end{figure}

In the conical refraction (CR) phenomenon \cite{kal2008,bel1978,ram1942,ber2007,bel1999}, a focused randomly polarized input Gaussian beam propagating along the optic axis of a biaxial crystal (BC) is transformed into a light ring, as shown in Fig.~\ref{fig1}, being this light ring most sharply resolved at the ring (focal) plane of the system. The ring radius, $R_0$, depends on the crystal's length, $L$, and its conicity, $\alpha$, through $R_0 = L \alpha$. For $R_0 \gg w_0$, where $w_0$ is the waist radius of the input beam, the light ring splits into two concentric bright rings separated by a dark region, known as Poggendorff dark ring. Furthermore, each pair of diagonally opposite points of the CR ring have orthogonal linear polarizations (see Fig.~\ref{fig1}(b)). The ordinary polarization corresponds to the point with polarization tangent to the CR ring and the extraordinary polarization appears at the opposite ring's point.
As the imaging plane is moved away from the ring plane defined at $Z = 0$, the light ring, see Fig.~\ref{fig2}(a), evolves into a more complex structure with secondary rings, see Figs.~\ref{fig2}(b) and \ref{fig2}(c). Then, at a certain distance $Z_{Raman} = \pm \sqrt{2} \frac{R_0}{w_0} z_R$, where $z_R$ is the Rayleigh range of the focused input beam, most of the light becomes concentrated at the center of the pattern in what is known as the Raman spots \cite{ram1942}, see Fig.~\ref{fig2}(d). Since the evolution of the CR ring is symmetric with respect to the ring plane, here we will consider only the region between the CR ring and the second Raman spot. 

In spite of being a relatively old phenomenon, only few articles have addressed CR in the nonlinear regime 
\cite{shi1969,sch1978,str1980,vel1980,kro2010,zol2011,pee2011}, being all of them centered in the study of second harmonic generation (SHG) processes. In most of these works \cite{shi1969,zol2011}, CR and its SH signal were generated in the same biaxial crystal. This configuration ensures a very compact set-up, but unfortunately the phase matching direction does not coincide, in general, with one of the the optic axis of the crystal. Therefore, only materials with very large nonlinearities are able to simultaneously generate SH and CR. An alternative study of SHG in combination with CR can be carried out by placing a nonlinear crystal (NLC) after a BC \cite{pee2011}. This configuration allows the adjustment of the phase matching condition for SHG in the NLC with the optic axis of the BC. In this case, a more efficient SHG process is expected. Our aim here is to study SHG in type I (LBO) and type II (KTP) NLCs of a beam transformed by the CR phenomenon. This configuration is also significantly interesting due to the influence of the wave-vector and polarization distribution of the CR beam over the SH process. We will analyze from an experimental and a theoretical point of view the resulting transverse intensity patterns for both types of SHG and we will compare the evolution resulting SH beam with that one the CR beam at fundamental frequency. 

\begin{figure}[]
\centerline{\includegraphics[width=1 \columnwidth]{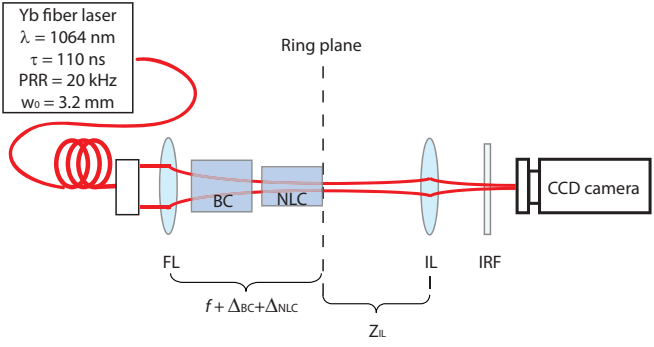}}
\caption{Experimental set-up. A randomly polarized input beam with a beam waist radius of $w_0~=~3.2\,\rm{mm}$ is obtained from an Yb fiber laser generating light pulses at $1064\,\rm{nm}$ with pulse duration $\tau = (110 \pm 10)\,\rm{ns}$ at $20\,\rm{kHz}$ repetition rate and up to $10\,\rm{W}$ of nominal power. This beam is focused by a lens (FL) of $400\,\rm{mm}$ focal length to a KGd(WO$_4$)$_2$ BC of length $L = 28\,\rm{mm}$ and conicity $\alpha = 17\,\rm{mrad}$, yielding $R_0 = 476\,\rm{\mu m}$. At the ring plane, we place the NLCs: LBO (type I, $d_{\rm{eff}}=0.668\,\rm{pm/V}$, $L_{\rm{LBO}}=10\,\rm{mm}$) and KTP (type II, $d_{\rm{eff}}=3.2598\,\rm{pm/V}$, $L_{\rm{KTP}}=8\,\rm{mm}$). The imaging lens IL projects different planes of the SHG propagated beams onto the CCD camera. The infrared filter (IRF) eliminates the radiation at the FH. $\Delta_{\rm{BC}}=L(1-1/n_{\rm{BC}})$ and $\Delta_{\rm{NLC}}=L_{\rm{NLC}}(1-1/n_{\rm{NLC}})$ are the longitudinal shift of the ring plane added by the BC and the NLC, respectively.}
\label{fig3}
\end{figure}

Fig.~\ref{fig3} shows in detail our experimental set-up. 
A randomly polarized input beam at $1064\,\rm{nm}$ is focused to a KGd(WO$_4$)$_2$ BC under CR conditions. At the ring plane, where the CR ring at fundamental harmonic (FH) appears, we place the NLCs oriented precisely under phase-matching conditions for optimal generation of SH. The focusing lens used ensures operation under plane wave approximation, i.e., $L_{\rm{NLC}} \approx z_R$ so that the NLC generates SH only from a unique transverse pattern of the CR beam. The length of the NLCs ($L_{\rm{LBO}}=10\,\rm{mm}$, $L_{\rm{KTP}}=8\,\rm{mm}$) is smaller than the distance of the Raman spot from the ring plane: $Z_{\rm{Raman}} \approx 166\,\rm{mm}$. Finally, an imaging lens (IL, with position Z$_{\rm{IL}}$) of $200\,\rm{mm}$ focal length projects the ring plane into the CCD camera. 

\begin{figure}[]
\centerline{\includegraphics[width=1 \columnwidth]{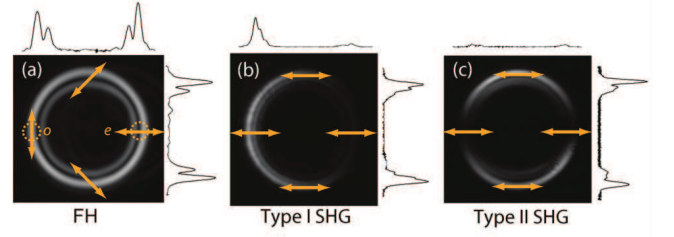}}
\caption{Patterns of the FH (a), type I (b) and type II SH (c) generated with the NLCs placed at the ring plane. Patterns were captured by using the lens IL, see Fig.~\ref{fig3}, to image the ring plane onto the CCD. Top and right insets are, respectively, the horizontal and vertical intensity profiles at the center of the images. Orange double arrows indicate the polarization plane.}
\label{fig4}
\end{figure}

Fig.~\ref{fig4}(b) and Fig.~\ref{fig4}(c) show the experimental SHG intensity patterns for type I and type II NLCs, respectively. We observe that in type I SHG, the transverse pattern consists of a light ring with a point of null intensity, resembling the pattern obtained in CR with linearly polarized beams \cite{kal2008}. In this case, however, the whole SH ring is linearly polarized with the polarization plane coinciding with the extraordinary polarization of the NLC. For type I NLC, SHG occurs in the form $oo \rightarrow e$. Thus, the point of the CR ring at the FH with polarization coinciding with the extraordinary mode of the NLC does not lead to SHG and, therefore, the resulting pattern in this case forms a crescent ring. With respect to type II SHG, we observe a light ring with two diagonally opposite points of null intensity. In this case, the ring is also linearly polarized coinciding with the extraordinary polarization of the NLC. Note that in type II SHG the doubling frequency process occurs through the channels $oe \rightarrow e$ and $eo \rightarrow e$. As a consequence, 
the two points of the CR ring at the FH with only ordinary or extraordinary polarization do not contribute to SH while the maximum SH intensity comes from those points of the FH ring with an equal contribution of ordinary and extraordinary polarizations.

\begin{figure}[t!]
\centerline{\includegraphics[width=1 \columnwidth]{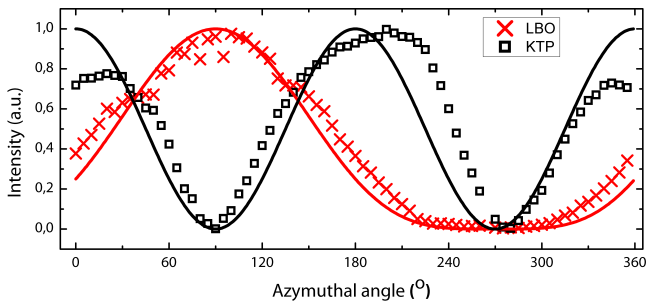}}
\caption{Azimuthal intensity distribution of the final patterns for type I (LBO) and type II (KTP) SHG. Symbols represent the experimental data, while solid lines are the corresponding analytical solutions from Eqs.~(\ref{eqtypeI}) and (\ref{eqtypeII}).}
\label{fig6}
\end{figure}

To obtain a quantitative description of the SH process we use the diffraction theory of CR derived by Belsky and Khapalyuk \cite{bel1978} and later on reformulated by Berry \cite{ber2007}, whose solution for the electric amplitude of the CR beam is:
\begin{equation}
\left|f \left(\xi, Z \right) \right| = \frac{e^{-\frac{\xi^2}{2 w_Z}}}{2^{5/4} w_Z^{3/4}} D_{\frac{1}{2}} \left( \frac{\sqrt{2} \xi}{\sqrt{w_Z}} \right), 
\label{eqaux} 
\end{equation}
where $w_{Z} = 1+ i Z$ and $D_{\frac{1}{2}}(x)$ is the parabolic cylinder (Weber) function \cite{bel1999}. $\xi \equiv \rho - \rho_0$ and $\eta \equiv \frac{Z}{Z_{R}}$ are the normalized radial and longitudinal components in cylindrical coordinates, respectively. Taking into account both the polarization and the intensity distribution of the FH given by Eq.~(\ref{eqaux}) and the the nature of the SHG processes ($oo\rightarrow e$ for type I; $oe\rightarrow e$ and $eo\rightarrow e$ for type II), it is straightforward to derive the corresponding analytical expressions for the SH intensity patterns from the CR ring:
\begin{eqnarray}
I_{\rm{Type\,I}} = I_{2 \omega_0} \frac{\left|f \left(\xi, \eta \right) \right|^4}{\rho_0^2} \cos^4 \left( \frac{\varphi + \phi_0}{2} \right), 
\label{eqtypeI} \\
I_{\rm{Type\,II}} = I_{2 \omega_0} \frac{\left|f \left(\xi, \eta \right) \right|^4}{\rho_0^2} \sin^2 \left( \varphi + \phi_0 \right), 
\label{eqtypeII}
\end{eqnarray}
where $\varphi$ indicates the point of the CR ring, $\phi_0$ is the mutual orientation between the planes of optic axes of the BC and the NLC and $\rho_0 \equiv \frac {R_0}{w_0}$ measures the ring radius in beam waists. $I_{2\omega_0}$ is the normalized intensity of the SH signal at the exit of a NLC for a Gaussian beam \cite{boyd2008}. In Fig.~\ref{fig6}(a) we plot the azimuthal intensity variations obtained experimentally (symbols) and the corresponding theoretical solutions (solid lines) for type I (red) and type II (black) NLCs. 

Note, in addition, that the inner Poggendorff ring is almost non visible in the SHG intensity patterns, see Figs.~\ref{fig4}(b) and \ref{fig4}(c). For the FH, it has been shown that the intensity of the input beam redistributes between the two Poggendorff rings in a ratio 3:1 (outer:inner) \cite{bel1978}. Since $I_{2 \omega_0} \propto I_{\omega_0}^2$, the two Poggendorff rings do not generate the same SH intensity signal, being the SHG outer ring much more intense than the inner one. 

\begin{figure}[]
\centerline{\includegraphics[width=1 \columnwidth]{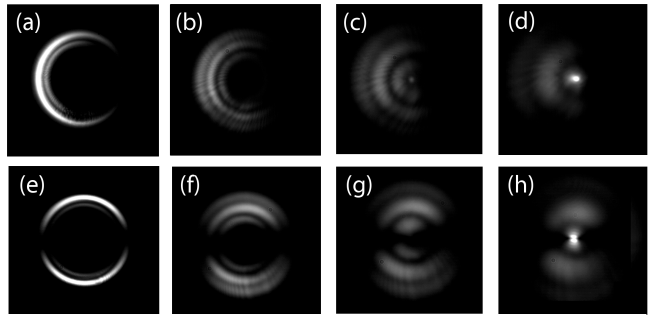}}
\caption{Evolution of the transverse intensity profile in type I (top row) and type II (bottom row) SHG when the NLCs are placed at the ring plane of the CR beam. The extraordinary polarization in the NLC was perpendicular to the plane of optic axes of the BC, i.e., $\phi_0=0^{\circ}$. We note that the Raman-like spots for second harmonic, see (d) and (h), have been observed on both sides from the ring plane.}
\label{fig7}
\end{figure}

Fig.~\ref{fig7} presents the evolution of the transverse intensity patterns of the SHG beams obtained by imaging different planes along the beam propagation. Comparing with Fig.~\ref{fig2}, one concludes that the frequency doubled waves are also conically refracted beams, being their evolution completely analogous to the FH. We have observed two focusing spots placed symmetrically from the ring plane of the SH signals resembling the Raman spots of CR. This behavior is expected since SHG is a nonlinear process that converts both the intensity and the phase of the incoming wavefront inside the NLC. 

In summary, we have reported SHG in type I and type II NLCs from an input beam refracted conically after passing along the optic axis of a biaxial crystal. This configuration allows aligning precisely the phase-matching direction of the NLC. For type I, the SH pattern at the ring plane forms a light ring with a point of null intensity, corresponding to the extraordinary polarization of the FH. In contrast, for type II SHG, the light ring possesses two dark points that correspond to the two points of the FH with only ordinary or extraordinary polarizations. We have provided a qualitative explanation of the SH intensity patterns in terms of the different channels that contribute to the SH signal and derived an analytical solution that is in good agreement with the experimental results. Finally, we have investigated the spatial evolution of the SH beams showing that they resemble conically refracted beams.  

The authors gratefully acknowledge financial support through Spanish MICINN contracts FIS2010-10004-E and FIS2011-23719, and the Catalan Government contract SGR2009-00347. A. T. acknowledges financial support through grant AP2010-2310 from the MICINN.

\end{document}